# Towards a Framework for a New Research Ecosystem

Roberto Savona[*,a], Cristina Maria Alberini[b], Lucia Alessi[c], Iacopo Baussano[d], Petros Dellaportas[e], Ranieri Guerra[f], Sean Khozin[h], Andrea Modena[h], Sergio Pecorelli[a], Guido Rasi[i], Paolo Daniele Siviero[j] and Roger M. Stein[k, l]

This draft: 12 December 2023

[*]Corresponding author. Roberto Savona, Dept. of Economics and Management, University of Brescia, C/da S. Chiara 50–25122, Brescia, Italy. roberto.savona@unibs.it

[a]University of Brescia; [b]Center for Neural Science, New York University; [c]European Commission–JRC; [d]International Agency for Research on Cancer (IARC/WHO), Early Detection, Prevention and Infections Branch, Lyon, France; [e]University College London & Athens University of Economics and Business; [f]National Academy of Medicine, Italy; [e]University of Mannheim; [h]Laboratory for Financial Engineering, MIT, [i]University of Rome Tor Vergata; [j]Farmindustria; [k]Stern School of Business, New York University; [l]NYU Center for Data Science.

# Towards a Framework for a New Research Ecosystem

This draft: 12 December 2023

Abstract: A major gap exists between the *conceptual* suggestion of how much a nation should invest in science, innovation, and technology, and the *practical* implementation of what is done. We identify 4 critical challenges that must be address in order to develop an environment conducive to collaboration across organizations and governments, while also preserving commercial rewards for investors and innovators, in order to move towards a new Research Ecosystem.

*"You may ask … How can I be a millionaire and pay no taxes …?"*

*"FIRST: Make a million dollars …"*

Steve Martin

# 1 Introduction[‡]

The global COVID-19 pandemic provided a "natural experiment" that demonstrated how international public-private collaborations can be not only *feasible* but also *effective* for addressing worldwide emergencies [1]. A coordinated, collaborative work approach was instrumental in securing social and economic pandemic resilience at an international level [2]. We learned how clearer and more coordinated scientific advice would reflect on more efficient policy decisions and public communication. We also "measured" the potential of public-private partnerships, which were essential in the battle against the pandemic.

Winston Churchill has been quoted[1] as admonishing that one should, "never let a good crisis go to waste." Churchill was commenting on the unique set of events that took place during and immediately following World War II that brought world leaders together to form the United Nations. The medical and scientific communities are now in a similar position, having been brought together and rallied to address a global health crisis, and having produced a number of effective treatments and prophylactics to control COVID-19.

Could a similar coordinated approach, post-crisis, be applied more generally, to help increase investments in science, innovation, and technology (SIT)?

Perhaps. But there remain significant challenges.

A significant challenge that hinders structured investments in SIT is often the absence of a pressing crisis that would justify transformative objectives. Furthermore, the high costs and risks of investing in transformative "long shots" – initiatives with sometimes unattractive individual investment profiles, but which produce outsized returns on success [8] – require discipline and patience to be successful, which can further complicate decisions to pursue these investments.

As a result, society as a whole has often missed potential opportunities to reap massive tangible and intangible benefits from long-term scientific research, due to a relatively shorter-term investment focus coupled with a lack of motivation to pursue unorthodox, often risky, approaches to supporting scientific progress.

---

[‡]We are grateful to Michael Spence and to a former government official for detailed comments and suggestions on earlier drafts. All errors are our own.

[1]Though Churchill does appear to have said this, it is not clear that he was the first; the provenance of the original quote is ambiguous.

Indeed, investments in visionary research and development (R&D) projects that hold the prospect of enormous social and economic value are often curtailed due to the empirical record of investments in which the returns on such projects can potentially take years or decades to emerge.

At the same time, the long-term and variegated benefits of basic scientific discoveries, though hard to predict *ex ante*, can have significant economic, environmental and social benefits across a wide array of applications and fields. Historically, many basic scientific discoveries have ultimately resulted in outsized impacts on social welfare, while conferring catalytic effects on other research and investment opportunities (and generating attractive returns for investors). In many cases, the future impact of a scientific discovery is only envisioned vaguely, if at all, at the time the related research is undertaken. Some recent examples include applications of CRISPR gene-editing technology [3], Tesla's leveraging of battery and solar technologies [4], and the production of mRNA vaccines for COVID-19 and other conditions [5].

Despite these proof cases, aggregate global government investments in basic and translational science is generally channeled towards supporting incremental progress, and undertaken largely without the benefit of substantive coordination at an international level [6]. Moreover, since the economic value that results from innovative projects is typically not fully appropriated by developers/inventors, the private sector tends to underinvest in ambitious large-scale scientific projects [7], even when they are characterized as having attractive payoffs for success, because of the corresponding high up-front development costs, high risk of failure, incompatibility with standard workflows, and long gestation periods [8].

While the returns on successful innovations can be extraordinarily attractive, the very high probability of failure and the long development time-horizon have made such investments attractive to only few specialized private investor classes, whose funding collectively represents only a small fraction of investment capital more broadly. Furthermore, due to regulatory uncertainties and the potential for future competing innovations, it can be uniquely challenging to estimate the commercial viability of long-term development projects accurately.

The 2015 *Report of the Scientific Advisory Board of the UN Secretary-General* [9] reported that many governments espoused the view that a target funding allocation of as little as 1% of global GDP for R&D was still too high. Nonetheless, the subset of countries with the strongest SIT systems actually invested as much 3.5% of national GDP in science, innovation and technology research.

As an alternative to one-off independent investments in SIT, strategies that encouraged international coordination, along with public-private collaboration that would combine public funding with additional private capital from investments by institutional investors, hold the potential to increase the scale and efficiency of R&D investment in SIT. The key to such strategies would be to create systems and incentives that would serve to catalyze investment in early-stage and translational R&D, by using the experiences of the COVID-19 pandemic response to inform more collaborative models.

In this short note, we outline 4 of the *key issues* that we have identified as critical challenges that must be address in order to develop an environment conducive to collaboration across organizations and governments, while also preserving commercial rewards for investors and innovators:

1. Increasing funding incentives through lower risk financing models;
2. Researcher collaboration through various physical and virtual cross-disciplinary laboratories that create both incentives and infrastructure for sharing novel research results at all stages of the scientific process
3. Demarcating specific roles and opportunities for governments and public institutions (GPI), and
4. Revising regulatory and legal frameworks for governance and control of research, including those for protecting and sharing rights to IP.

While these challenges are prevalent in various organizational and institutional settings more generally, they also present particular opportunities for global research efforts in the scientific domain due to their unique characteristics and nuances.

Rather than offering definitive solutions, our intention in what follows is to present a positive (i.e., non-normative) perspective on these challenges, and to outline a basic set of signposts to stimulate dialogue among experts in various fields including science, medicine, computer science, finance, law, policy, and academia. Our goal is to encourage global discussion, debate, and ultimate action that can contribute to the design of a new research ecosystem.

## 2  Four Challenges

### 2.1  *Challenge #1: Accelerating Funding*

Although funding is often discussed as the primary challenge to precipitating long-shot investments, it is ironically the challenge for which we already seem to have many of the most well-developed solutions.

We describe many investment in SIT projects as being "long shots" in the sense of Hull, Stein and Lo[8]. Long shot investments are characterized by (a) a very low probability of success; (b) long investment horizons; (c) substantial upfront capital requirements; but which also enjoy (d) outsized commercial returns upon success. Empirically, the majority of retail and institutional investors prefer more traditional, less risky investments that enjoy shorter return horizons and lower expected returns, with much lower outcome uncertainty.

Despite these seemingly less attractive features, long shot investments can often offer attractive returns when successful. Science-focused investment vehicles can potentially mitigate this long shot problem by employing results from modern portfolio theory, which provides guidance on structuring *portfolios* of investments, such that the volatility (risk) of the portfolio is reduced, while still maintaining attractive return profiles, relative to the individual investments in the portfolio. Properly structured portfolios of long-shots can greatly reduce the risk to investors, while still maintaining a comparable return.

Recent work by financial economists has demonstrated that risk pooling structures can serve as the basis for effective funding vehicles that provide longer-term capital for research, while also delivering *market-rate* returns for investors. Said differently, such structures can achieve substantial social and scientific impact, without requiring that investors give up investment returns.

Most notably, a new class of securities, called Research-Backed Obligations (RBOs) – namely, debt and equity securities backed by the pool of underlying drug assets issued by 'mega-funds' to raise capital and finance the development of pipeline drugs in its portfolio –, are designed to fund portfolios of pooled long-shot research investments in candidate medical therapies for cancer and rare genetic diseases by taking advantage of portfolio diversification to issue high-quality (and thus lower cost) portfolio-level debt [10], [11], [12]. Similarly structured financial vehicles could play a key role in increasing investment in other areas of scientific research. Importantly, these new structures *do not require that investors trade return for impacts*. Rather than forcing investors to give up investment return to achieve social impact, RBOs can offer attractive risk/return profiles, while also generating impact.

However, not all R&D challenges are suitable for vehicles such as RBOs. For example, the current state-of-the-art in Alzheimer's research, along with the dearth of viable AD (Alzheimer Disease) therapy projects appears to make funding research in that area via an RBO structure difficult due to an extremely high probability of failure and a lack of appropriate diversification options at the present time [18].

Thus, while the RBO approach has wide applicability in a many scientific domains, and has demonstrated some early successes in biomedical research, there remain other areas which are not suitable to such an approach. In such settings, given the enormous scale of investment required and the exceedingly low *ex ante* probability of success, it may be that only a well-funded socially motivated entity, like a government agency, can ensure sufficient capital and investment discipline to fund research. [8]

Between these two different sources of funding – public vs. private – there exists a middle ground, in which projects representing "ultra-long-shot" research may still be attractive to private investors, but may not necessarily require public funding. For such investments, hybrid financing approaches have also been introduced in the economics literature. For example, an approach suggested by [13], may offer tools that support the creation of vehicles to fund "ultra-long-shot" R&D projects by combining investments in both high- and low-risk research projects, along with investments in traditional low-risk financial instruments such as government bonds (albeit with potentially lower expected portfolio returns).

Under such a hybrid strategy, a portfolio manager follows a *search and 'wait or invest'* approach: projects are ordered according to a specific scoring formula, defined by the portfolio mandate or investment policy (potentially augmented with a *strategic* consideration of the *knock-on contribution to other or future portfolio projects* of the research). These investment strategies require that (a) investment decisions be interconnected across time, geography, and research domains; and (b) portfolios are structured into sub-portfolios that differentiate between connected *lead* (riskier) projects and *backup* (less risky) projects that serve to reserve capital for future development. In some settings this coupling approach may reduce the risk that lead projects fail due to a lack of interim funding. Of course, the expected return on a sub-portfolio of less risky projects is much lower than that of a sub-portfolio of riskier projects, so the overall portfolio return will necessarily be lower than that of a "pure-play" investment in the risky projects, but may nonetheless offer returns that are consistent with other investments in the market with similar risk profiles.

Hybrid approaches, such as these, obviously require detailed quantitative analyses to ensure that the reduction in the expected portfolio volatility is sufficient to offset the lower expected return of the backup projects.

Regardless of the structure of these pooled investments, the overriding design principle is that, if the goal of the resulting investment product is to target *private investors*, it must offer *market appropriate risk-return profiles* in addition to achieving any additional knock-on altruistic social goals.[2] Said differently, the pool of long-shot investors must be extended beyond only those investors who put high value on achieving social objectives (by willingly giving up return in exchange for doing "good"), to those investors who don't mind supporting social objectives, but who are primarily motivated to invest for financial returns. This second group is much, much larger than the first.

An example of a mandate of an SBF (Science-Based Fund), such as the one shown in Figure 1, would be to concentrate on one or more broad scientific objectives, and could span a number of phases of R&D, from basic research through translation of promising discoveries, culminating in commercialization.

Such a vehicle might be sponsored in a number or ways. Examples include public sector coalitions of government agencies, private sector coalitions of venture capitalists and other institutional investment professionals, and public-private partnerships, which we discuss in the next section.

## 2.2  *Challenge #2: Research collaborations*

Operationalizing such large-scale development is, at best, challenging, since the governance required to align incentives for R&D translation can be complex, and also involve substantial coordination and participation from diverse stakeholders with varying priorities. Furthermore, governance, which we discuss in Section 2.4 becomes even more complicated in settings in which public and private funding is used to finance these types of projects.

In this subsection, we discuss the mechanics and challenges of *coordinating development, communication and funding* (this section); and then discuss a number of the options and the potential roles that government agencies and NGOs can play in moving a global research ecosystem forward (next section).

---

[2] If this is not possible (as in the case of Alzheimer's therapies in an RBO portfolio), third party support from governments, NGOs or foundations may be required to bring fund the projects or to bring the risk-return profile back in line with market levels, or both.

A number of researchers (e.g., [14]), have reported empirical evidence suggesting sharp declines in research productivity over the first two decades of the 21$^{st}$ century. There appear to be many reasons for this. However, from a practical perspective, two pervasive logistical challenges have emerged as critical hurdles:

1) difficulties in developing and extending promising early-stage ideas to translate them into real-world solutions to address key social problems (acceleration to innovation and translation), which implies,
2) difficulties in securing long-term "patient" funding (long-term financing).

In the absence of more capital from private investors, promising and even disruptive projects can remain idle indefinitely because they cannot generate enough support among academic laboratories, start-up firms or governments to gain financial sponsorship through traditional funding sources.

One institutional response to this dearth of funding has been the creation of a number of government and academic incubators that support non-profit-type start-up organizations [15] by providing funding for research projects, facilities and other services.

In our earlier discussion of accelerating funding, we highlighted the advent of recent alternative funding structures, motivated by results in portfolio-theory and other areas of finance, that can supplement or substitute for such incubators.

However, such structures, in isolation, are only part of the solution. Regardless of *how* such investments are structured, delivering on such broad investment mandates and pooling multiple projects across scientific disciplines or organizations requires a robust coordination mechanism that minimizes potential agency costs between the principals (investors in and sponsors of the investment vehicles) and the agents (the managers of the vehicles).

In principle, for example, the design of such a mechanism might contemplate that the principals (investors or sponsors) define the very general scientific mandate of the fund, while the agents who manage the fund coordinate management of the fund to achieve that mandate. In practice, of course, such governance mechanisms are nontrivial to negotiate and implement, particularly at an international level.

How might such a coalition of principals and agents' function? Is forming such a coalition even feasible and desirable? Would it solve agency problems?

Although these concepts have yet to be tested at scale for scientific research, there are already encouraging proof-cases of similar examples of collaborative projects/approaches in other disciplines. For example, the Polymath project [16] demonstrated that an open, collaborative group of mathematicians can participate in a shared platform to solve open research mathematical problems. As another example, open laboratory notebooks, in which both academic and industrial chemistry scientists recorded their work in standard online form, have also demonstrated that researchers can develop new scientific breakthroughs in a synergistic manner [17].

One overarching goal of scientific collaboration networks focused on specific areas of basic science research could be to form what we term *virtual labs*: distributed teams of researchers and technicians who work on various components of a problem remotely, and then integrate their work through virtual or in-person workshops. The historical barriers to forming and operating such organizational structures are now greatly diminished, as the COVID-19 lockdown experience demonstrated. Forming distributed working groups and teams that operate both synchronously and asynchronously has become much more practicable by virtue of advances in distributed ledgers, video conferencing technologies, cloud storage and computing infrastructure and emerging virtual reality communication tools.[3]

Mission-driven discovery can also be accelerated and supported through the active use of meta-knowledge management and analysis tools that efficiently leverage data and expertise. Text mining and clustering, automatic linkage of patent-article network connections, federated learning algorithms and large-language models can help distill the substance and direction of large volumes of disparate research, while also preserving privacy in many cases.

Though overreliance on such tools may subvert more fulsome analysis of relevant information, and despite their current limitations and biases (e.g., [18]), these advances will continue to evolve. Despite their limitations, the hold the potential to provide unprecedented reach to researchers globally. This highlights the importance of, and the natural role for, data science researchers and computer scientists, whose expertise and research can be leveraged to identify overlooked exploration or recapitulation and to surface new research connections with novel paths that cross discipline-specific knowledge boundaries. [19]

A significant amount of scientific innovation originates from academia, but it is often concentrated in specific locations or disciplines. Universities, for instance, can have a diverse range of experts and knowledge even within a single department or campus, but this rich knowledge-base is often isolated from similar groups in other disciplines. It is often even difficult for researchers within the same institution or city to collaborate with researchers in different disciplines.

Properly structured virtual labs may offer the potential to reduce barriers to communication through technological solutions that aid in forming research communities in which research teams can discover others working in related, adjacent, or even seemingly unrelated areas, and who may have the skills and expertise to fill key gaps in a research agenda. This flexibility is particularly important in emerging fields in which the need for extra-disciplinary expertise may become obvious only as research progresses. For example, so called "Science of Science" studies that bring into relief past relational structures between scientists or scientific disciplines, could provide insights into the genesis of scientific discovery [20] or even be used to target a specific scientific objective.

---

[3] Parenthetically, these technologies have resulted for the most part from early investments in basic research.

## 2.3 Challenge #3: Structuring the Support of Governments and Public Institutions (GPI)

Governments and public institutions can play a number of key roles to increase the feasibility of creating research ecosystems and to encourage their realization.

Some of these roles include:

1. *Independent Project Evaluation.* One of the advantages of integrated research agendas and systems is that they offer the potential to scale up investment in R&D by channeling private capital, including capital from financial investors who do not have specific expertise in a particular research field. Unlike traditional venture capitalists, such investors are often not in a position to evaluate the merits or technical detail of a specific project. Public Institutions, in the form of new agencies, can develop standards for independent projects' ratings or rankings relative to their expected marginal contribution to enhancing scientific knowledge and their environmental and societal impacts, while simultaneously providing regulatory clarity on the path to future commercialization.

   Practically, it will be useful to further segment evaluation functions into sets of very broad objectives. For example, it may be advantageous to use different rating methodologies and metrics to evaluate proposed initiatives depending on whether their objective is to (e.g.):

   - advance basic knowledge in some area – with no other specific goal or with goals that are likely unknown at the time of the research. (Such initiatives may be of little interest to private investors); or
   - translate research results into commercial applications in the form of long-term risky investments that do have a goal, like reducing or curing cancer. (Such initiatives often involve intensively top private investors).
   - Project Pooling. In cases where public-private collaborations do make sense on both sides, members of the public sector can facilitate these collaborations, by providing incentives for pooling projects that advance science in those fields where industry investment is insufficient or non-existent.
   In this context, GPIs can leverage their natural authority to:
     a. convene participants and act as clearing houses for information;
     b. create incentive programs and tax relief;
     c. make direct investment (as discussed in the next point);

   and so forth.

2. *Minimum return guarantee.* If a pool of projects can be made large and diversified enough, it is reasonable to expect that investments in them will be remunerated, with a potentially extraordinary upside potential. However, determining the value of a share in such an investment can be difficult due to the very high variance of and uncertainty surrounding future cashflows. It is often hard enough to estimate future cash flows that will accrue for translating research that has already been developed (but not yet commercialized), and far more difficult still to do so for earlier stage research and IP.

In this context, GPIs could reduce investor uncertainty through some form of downside protection, either by providing direct guarantees, thereby reducing risk for private investments (by covering the first-loss piece, for example); by purchasing shares of a Science-Based Fund or designated classes of RBO securities that would stand in a first-loss positioning in the capital structure (perhaps doing so at below market rates if needed),[21] or by providing purchase guarantees for the resulting products that are brought to market. It can be shown that investments in the form of backstop guarantees, for example, have the potential to amplify and multiply government investments by attracting traditionally reluctant private investors to participate [21]. The Israel Life Sciences Fund provides one real-world proof case of such a vehicle. This fund was sponsored by the Government of Israel but owned and managed by the private sector. The government of Israel invested alongside institutional investors, but received a subordinated return, as a means to attract investment.[22]

In summary, we believe that coordinated public-private efforts are necessary for driving investment in some kinds of transformative SIT initiatives. These efforts may take the form of independent projects, collaborative public-private partnerships, or synergistic yet arms-length arrangements, depending on the specific circumstances.

## 2.4  Challenge #4: Governance, Control of Research and IP

A research ecosystem along the lines that we discussed in Section 2.2 and 2.3 envisions new organizational architectures in which researchers from universities, industry, government join virtual labs and interact physically and electronically, supported in both modes by a shared analytic infrastructure (e.g., graph databases, federated data stores, data science tools, Research and Technology Infrastructures[4], etc.). The research projects that emerge from these collaborations will need to be evaluated and, possibly, triaged. Some projects may be well suited to being pooled in some form such as within project portfolios by a Public Agency, as outlined in Section 2.3, and some may be better suited to straight private sector investing. In either case, pools of assets can be financed more efficiently using structures such as RBOs, megafunds or other Science-Based Funds, alongside government efforts.

Of course, this can only work if the various participants and stakeholders have sufficient comfort, through rigorous rules of engagement that mitigate abuse and waste and enjoy sufficient incentives, both commercial and career-related.

One step in this direction would be to ensure that research projects, and the associated cashflows to and from them, are clearly accounted for and reported as assets on the funding vehicle's P&L (Profit and Loss statement) and balance sheets (as either real or intangible assets, respectively, depending on the nature of the item).

---

[4]   E.g.,   https://research-and-innovation.ec.europa.eu/strategy/strategy-2020-2024/our-digital-future/european-research-infrastructures_en

Proper governance and incentives, as well as the allocation of economic and ownership rights, are crucial for the successful development of new intellectual property through joint efforts among participants. In most cases, it is essential for individual contributors and organizations to effectively manage their sharing of information through participation in collaborative agreements that address both scientific and business goals.

Historically, it was often the case that only a government or NGO could provide incentives and constraints to help ensure that the contributions of each constituent in an entire portfolio, including the spill-over benefits of foundational research, is enumerated and factored into private sector investment decisions. However, with the advent of distributed ledgers, homomorphic encryption, and privacy-preserving machine learning techniques, controlling the usage and auditing the provenance of specific contributions of IP and data may now be feasible in a manner not possible in even the last decade.

Regulatory bodies will also enjoy windfall opportunities to rethink the requirements and restrictions they have already put in place for different forms of research and subsequent commercialization. Importantly, although these protocols drive the frameworks used for translational science in most domains, most of these regulatory frameworks are still relatively new. For example, many features of current drug development regulation are only decades old. In many cases, important portions of the regulatory bulwarks in place today arose from amalgams of situational and pragmatic responses to public health crises or public outcry, rather than necessarily having benefited from a calculated well-crafted, internally consistent design and vision.

To be clear, though, while we do see an active, and essential, role for governments and NGOs, by virtue of their budgetary and regulatory authority, *we are not proposing either that the public sector "federalize" research, or that entrepreneurs be curtailed from realizing the benefits of their creativity and labor*. Rather, we are suggesting that researchers from industry, government and the private sector would benefit from controlled, sometimes coordinated, appeals for more eyes or minds on specific aspects of interesting problems along with assurances that their contributions will be recognized and rewarded. It remains an open (and empirical) question whether this approach would be best achieved through consortia, jointly funded research projects, a marketplace for expertise and data, or some other form.

However, there is cause for optimism. The recent COVID-19 pandemic (again) provides a real-world example of successful public-private organizational architecture. The Accelerating COVID-19 Therapeutic Interventions and Vaccines (ACTIV) platform created in the United States enabled biopharmaceutical firms to develop COVID-19 vaccines at a much faster rate than traditional approaches. Indeed, this unprecedented partnership brought together [23]: (i) impressive levels of public funding; (ii) shared competences from across an 'ecosystem' of universities, government entities (Biomedical Advanced Research and Development Authority, National Institutes of Health, Food and Drug Administration), biotech and big pharma companies, and even the U.S. Army; and (iii) a small, efficient and politically-independent unified governance structure to take decisions quickly. This governmental intervention was essential to vaccine development, first, by providing massive support to private enterprise, second, by mitigating scientific as well as manufacturing and market risks related to possibly low demand [24].

It will also be important to recalibrate research incentives for both government agencies and academic institutions. It is natural for research centers to push back on collaborations that require coordination with other institutions because many researchers value "autonomy" highly, both for intellectual reasons and political and organizational ones.

Furthermore, creating incentives that encourage interorganizational collaboration can be especially challenging in academia, where data sharing is often inhibited by virtue of "publish or perish" merit systems at many academic institutions. This reward structure can overshadow many other motives. This incentive structure can also skew research quality and follow-through, since researchers may rewarded much more for producing "high quality" publications than for facilitating pragmatic applications of their scientific output. [25]

Universities and other academic and commercial research organizations that choose to participate in collaborative research initiatives and virtual labs will have the opportunity, and the imperative, to rethink, at a fundamental level, how collaborative research is recognized and rewarded at their institutions, as well as the degree to which translation of research ideas is weighed alongside innovation and breakthroughs when researchers are evaluated at different points in the research lifecycle.

## 3  Operationalizing a Research Ecosystem

The purpose of this article is to outline what we have identified as key impediments to the formation of multi-disciplinary pan-geographic collaboration projects to address critical and, in some cases, existential societal challenges. We have specifically not attempted to propose omnibus solutions to any of these. Such solutions require many more eyes and ideas that we can bring to bear in a single article, and will sometimes demand resources and legal frameworks which we have no standing to allocate.

However, there are things that can be done to start this process. For example, one way to begin such efforts is to create a research ecosystem architecture (see Figure 2).

As an initial, much less demanding foray into such planning and discussions, it is reasonable for smaller groups of researchers and other participants to undertake a set of pilot studies to explore what is currently feasible for addressing the four challenges we have outlined in this article. Such pilots would include experts from various relevant parties – academic, industrial, government and finance – with the goal of rigorously exploring – and perhaps competing – theoretical, technical, and institutional frameworks for the addressing specific aspects of the challenges we outlined earlier.

Because such an approach would require new contexts, such groups could be made more productive and agile by creating sandbox-based processes that would permit them to experiment in live environments, but with safeguards in place to ensure that such experiments did not result in unintended consequences in society and the economy more broadly.

It seems both productive and feasible to consider exploring several broad sets of questions in parallel. For example, different pilot groups might undertake topics relating to:

- Marketplace of Ideas and translational IP: The mandate of this pilot would be to realize a decentralized market where innovators would be able to sell ideas to entrepreneurs and institutional investors using financial structures and instruments whose profitability was contingent on those ideas. (challenge #1 and #4). Existing examples are EIC Marketplace (https://eic.ec.europa.eu/eic-communities/eic-marketplace_en), InvestEU Portal (https://ec.europa.eu/investeuportal/desktop/en/index.html), Innovation Radar (http://innovation-radar.ec.europa.eu), etc.
- Research Collaboration Platform: The mandate of this pilot would be to create a framework for building "knowledge maps" that would allow researchers across disciplines to more easily understand how innovations outside of their core field might be valuable to achieving a specific scientific goal. This would naturally precipitate scientific collaborations across researchers and ideas (challenge #2).
- Regulation and Supervision: The mandate of this pilot would be to draft a formal set of feasible options, both within current frameworks that could be achieved by introducing or modifying regulation and laws. This prolog would also enumerate and identify the roles and actors required to effectuate a research ecosystem supervision mechanism (challenge #3 and #4).

Table 1 summarizes this example of the different pilots and how they fit together as part of the ecosystem. In addition, it gives examples of how each main actor (University Presidents, Research Labs, Researchers, Investment Funds, Funding Agencies, Technologists, Governments) might participate, and how they should interact with each other within the main challenges (Market for Ideas, Research Collaboration, Regulation & Supervision).

Again, our intention here is not to set out a prescriptive formula for achieving a research ecosystem. Rather it is to stimulate dialogue among experts in various fields, encouraging global action and contributing to the design of a new research ecosystem. In that regard, we take the position that local stake-holders in their own of geographies and institutions, will attenuate the scope and scale of a pilot in the manner appropriate to their geography, institution and discipline.

# 4 Conclusion

At the end of 2020, UNESCO estimated that global spending in R&D normalized by GPD (R&D/GDP) was roughly 1.9% worldwide. If instead, all countries were to invest on par with the most advanced countries recorded, i.e., at 3.5% of GDP, then based on global GDP at the end of 2021 (around 96 trillion USD) nations would need to close an R&D funding gap of around 1.5 trillion USD. However, even if there were the will to do so, it is still unclear, without more robust coordination, whether the 3.5% observed is the right target, or is too low (or too high); and whether the additional capital would be deployed in an even moderately efficient manner.

In the past half century, a number of rigorous theoretical results have emerged in the economics literature, and provided explanations for *why* investments in novel, ground-breaking scientific projects drive increases in overall welfare [26]-[27]. However, transitioning from normative theoretical results to concrete operational measures is not straightforward, since the trajectory from constellations of often disparate discoveries and theories to the explicit economic growth they produce is complex, non-linear, multidimensional and slow. Furthermore, a major gap exists between the *conceptual* suggestion of how much a nation "should" invest in core R&D, and the *practical* implementation of what is done. (Consider again the gap between the 3.5% of GDP invested in SIT by *the most advanced* countries vs. the *average* of only 1.9% across national governments overall [9]).

One key reason for this gap is the difficulty that investors (and the public sector) have in trading off short-term costs against long-term planning and investing benefits.

A related issue, known as 'the tragedy of the horizon'[28], has already been identified as a gating factor in the context of climate change, where the devastating consequences of policy and investment decisions made today will, tragically, not be felt by society until times that extend well beyond the horizons normally considered by investors and politicians. In order to reorient investors, more work must be done to articulate the financial risks that are posed by climate change and other SDG (Social Development Goals) related shortfalls. In parallel, we need clearer articulation of the opportunities investing in such initiatives can provide.

In this short note, we have offered a coarse roadmap to some of the issues that we see as critical, and which must be driven to consensus in full or in part to motivate a more aggressive and fulsome scientific and financial transformation and redeployment of resources.

Our roadmap only provides of a guide to the *hazards* in this rough terrain, rather than a *course* through them.

Nonetheless, if these issues can be resolved, and initiatives of the sort we have sketched here were adopted on a larger multinational and multidisciplinary scale, a massive science-oriented portfolio allocation could potentially open the door to new ways of research collaboration.

Achieving such scale will require that private investors are (financially) motivated to channel their capital towards science-based funding by investing in financial vehicles that can achieve adequate (i.e., market-level) risk-return profiles while also providing positive societal impacts. While these financial issues and structures appear to be relatively easier to address, compared to those of governance, and so forth, this financial structuring itself is by no means trivial.

For policymakers in government and at academic institutions, the message is similarly clear: collaborative thinking and planning involving industry, academic and government decision makers is a precursor to developing regulatory, policy and academic incentives that enable such broad investment vehicles to form and operate at scale.

However, with these vehicles in place, capital from private sector investors will naturally flow towards funding scientific research in order to achieve attractive risk/return profiles, while also providing asset-class diversification. Such investment, in concert with public allocations, may well enable public-private partnerships that would encourage investment at levels that today appear practically viable for only the most advanced economies.

Coming full circle then, we conclude by revising the joke at the beginning of this article to reflect our conclusions about the current state of play for the four challenges we discused:

> *"How can we create an infrastructure that solves the problems of organizing large-scale scientific research communities and then fund it?*
>
> *FIRST: Create an infrastructure that solves the problems of organizing large-scale scientific research communities …"*

*Table 1: Example high-level plan for topics and actors in Research Ecosystem Implementation Working Pilots*

|  | Market of Ideas | Research Collaboration | Regulation & Supervision |
|---|---|---|---|
| **University Presidents** | IP agreements: joint ownership and foreground. Forming special independent vehicles for selling ideas in the market | Forming and participating in domain-specific academic research consortia | Forming stakeholder groups within the Market of Ideas Board and contribute to define market regulation protocol. Interact within regulatory sandbox platforms (Governments) |
| **Research Labs** | | Forming Consortia of Research Labs | |
| **Researchers** | Create and test framework for Virtual Labs for pilot working research projects (perhaps selected and evaluated by Governments) | Propose requirements for and structure of Virtual Labs, potentially forming small scale proof-cases. | |
| **Investment Funds** | Structure and launch investment vehicles such as megafunds or Science-Based Funds, perhaps through private-public initiative | Networking with solicited and unsolicited researchers (meetings, on-line platforms, open contests) to detect new forms of research combinations and start with new curiosity- or mission-driven projects | |
| **Funding Agencies** | Provide funding for Market of Ideas infrastructure | Provide funding for realizing Virtual Labs. Defining mission-oriented projects. Provide funding and scientific coordination of meetings and activities | |
| **Governments** | Draft feasible mechanisms for providing various forms of downside protection for investors (direct guarantees, direct investments, etc.) | Providing independent project evaluation. Realizing project pooling. Designing appropriate IP agreements | Design potential structures for regulatory sandbox platforms, including appropriate safeguards to contain inefficiencies and maintain the overall safety and soundness of the research ecosystem |

*Figure 1: Science-Based Fund*

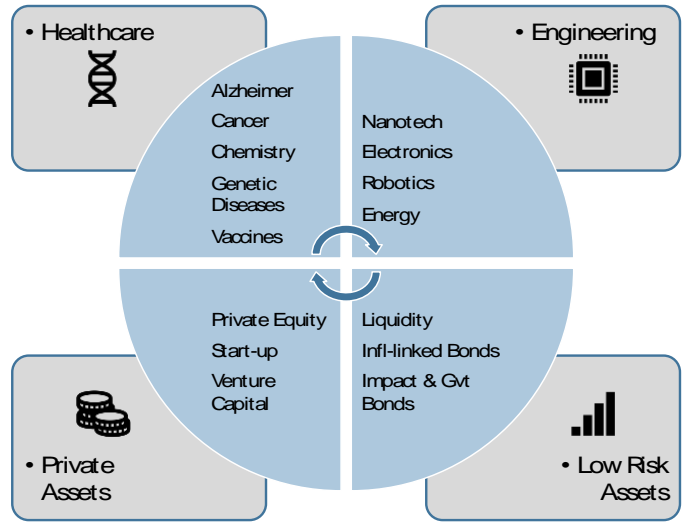

Note: The Figure depicts the organizational structure of a Science-Based Fund. Investments are diversified between Private Assets (Pure and Secondary Market of successful Venture Capital, namely Private Equity and Start-up) and Low Risk Assets (investments in liquid assets to provide *interim* returns to the fund and reduce the overall portfolio risk). Private Assets are investments in healthcare and engineering sectors, spanning from Alzheimer, Cancer, Chemistry, Genetic Diseases, Vaccines, on the one hand, to Nanotechnology, Electronics, Robotics, Energy, on the other. High and low risky research projects are combined with low risky financial instruments under a cross-funding approach in which returns from financial vehicles and backup (low risky) investments together with rotational portfolio reallocations are used to successfully fund lead (high risky) projects.

*Figure 2: A Hypothetical Research Ecosystem Architecture*

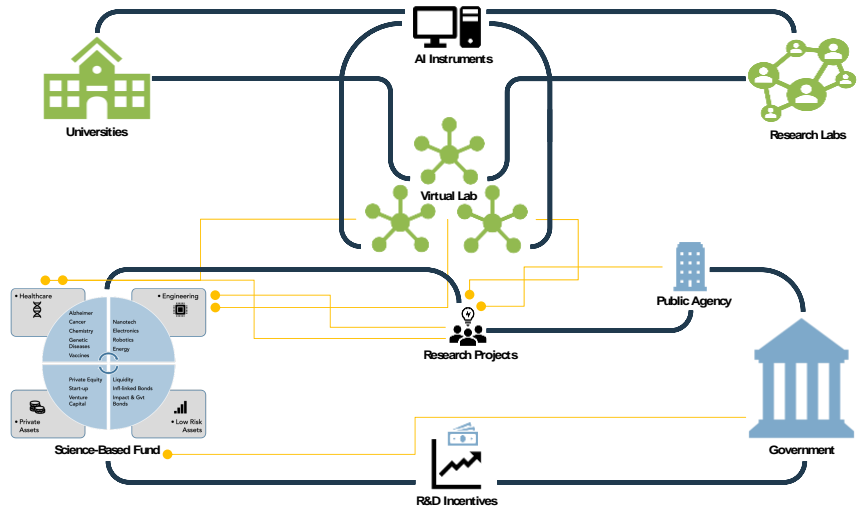

Note: The Figure shows a hypothetical Research Ecosystem Architecture with the 3 main actors: (1) Universities & Research Labs; (2) Science-Based Fund; (3) Government & Public Agency. Universities & Research Labs realize domain-specific academic research consortia and through advanced technology (AI Instruments) detect and cluster research ideas thereby forming Virtual Labs. Research Projects come from Virtual Labs, are evaluated by a Public Agency and incentivized (R&D Incentives) by the Government. S-Based Funds invest in Research Projects, benefit from government incentives and collaborate with the Virtual Labs. A possible action plan to operationalize this Research Ecosystem Architecture is reported in Table 1.